\documentclass[aps, prl,twocolumn, showpacs,superscriptaddress,10pt,floatfix]{revtex4-2}

\usepackage{graphicx}% Include figure files
\usepackage{dcolumn}% Align table columns on decimal point
\usepackage{bm}% bold math
\usepackage{hyperref}% add hypertext capabilities
\usepackage{xcolor}
\usepackage{mathrsfs}
\usepackage{physics}
\usepackage[page]{appendix}

\begin{document}

%\preprint{APS/123-QED}

\title{Operator Delocalization in Quantum Networks }

\author{Joonho Kim}
\email{joonhokim@ias.edu}
\affiliation{Institute for Advanced Study, Princeton, NJ 08540, USA}

\author{Jeff Murugan}
\email{jeff.murugan@uct.ac.za}
\affiliation{The Laboratory for Quantum Gravity \& Strings, Department of Mathematics and Applied Mathematics, University of Cape Town, South Africa}

\author{Jan Olle}
\email{jolle@ifae.es}
\affiliation{Institut de F\'{ı}sica d’Altes Energies (IFAE) \\ 
The Barcelona Institute of Science and Technology (BIST) Campus UAB, 08193 Bellaterra, Barcelona, Spain}

\author{Dario Rosa}
\email{dario\_rosa@ibs.re.kr}
\affiliation{Center for Theoretical Physics of Complex Systems, Institute for Basic Science (IBS), Daejeon - 34126, Korea}

% \author{Matteo Carrega$^{1}$, Joonho Kim$^{2}$, Jeff Murugan$^{3}$, Jan Olle$^{4}$ \& Dario Rosa$^{5}$}
% \affiliation{
%  $^{1}$NEST, Istituto Nanoscienze-CNR and Scuola Normale Superiore, I-56127 Pisa, Italy\\
%  $^{2}$ Institute for Advanced Study, Princeton, NJ 08540, USA\\
%  $^{3}$The Laboratory for Quantum Gravity \& Strings, Department of Mathematics and Applied Mathematics, University of Cape Town, South Africa\\
%  $^{4}$Institut de F\'{ı}sica d’Altes Energies (IFAE)
% The Barcelona Institute of Science and Technology (BIST) Campus UAB, 08193 Bellaterra (Barcelona) Spain\\
% $^{5}$Center for Theoretical Physics of Complex Systems,
% Institute for Basic Science (IBS),
% Daejeon - 34126, Korea}

\date{\today}% It is always \today, today,
             %  but any date may be explicitly specified

\begin{abstract}
%Motivated by recent developments in the theory of quantum batteries, 
We investigate  the delocalization of operators in non-chaotic quantum systems whose interactions are encoded in an underlying graph or network. In particular, we study how fast operators of different sizes delocalize as the network connectivity is varied. We argue that these delocalization properties are well captured by Krylov complexity and show, numerically, that efficient delocalization of large operators can only happen within sufficiently connected network topologies. Finally, we demonstrate how this can be used to furnish a deeper understanding of the quantum charging advantage of a class of SYK-like quantum batteries.

% OLD VERSION BELOW
% Motivated by recent developments in the theory of quantum batteries - small quantum systems with non-degenerate energy levels from which energy can be reversibly extracted, or deposited into, through cyclic unitary operations - we study  the delocalization of operators in many-body, non-chaotic, quantum systems whose interactions are encoded in an underlying graph. We show numerically, largely through use of a recently-introduced resource of Krylov-complexity, that efficient operator delocalization in a broad spectrum of quantum networks requires two essential ingredients:(i) a sufficiently connected network topology and (ii) operators of large enough size to exploit the graph topology. We show how this can be used to furnish a deeper understanding of the recently noted quantum charging advantage of a class of SYK-like quantum batteries.    
\end{abstract}

\maketitle

%\tableofcontents

%%%%%%%%%%%%%%%%%%%%%%%%%%%%%%%%%%%%%%%%%%%%%%%%%%%%%%%%%%%%%%%%%%
%\section{\label{sec:intro}Introduction}
%%%%%%%%%%%%%%%%%%%%%%%%%%%%%%%%%%%%%%%%%%%%%%%%%%%%%%%%%%%%%%%%%

{\it I Introduction.---} The conjecture that black holes are the fastest scramblers of information in nature \cite{Sekino_2008} has precipitated a renewed interest into questions of thermalization and ergodicity in quantum systems \cite{rigol_review_ETH}, and ushered in a new era of collaboration between seemingly disparate fields like high energy theory, condensed matter physics and quantum information. In this regard, one particularly important development in the past five years has been the emergence of the Sachdev-Ye-Kitaev (SYK) model \cite{Maldacena:2016hyu},
\begin{eqnarray}
    \hat{H}_{\mathrm{SYK}}^{(q)} = \mathrm{i}^{q/2} \sum_{i_{1} < \cdots <  i_{q}} J_{i_{1}  \cdots   i_{q}}\hat{\gamma}^{i_{1}}\cdots \hat{\gamma}^{i_{q}}\,,
\end{eqnarray}
of disordered Majorana fermions as a canonical framework to study questions from the information-loss paradox in (low-dimensional) quantum gravity to the physics of spin-glasses. The SYK model in turn has led to the development of a host of new (or, sometimes, forgotten) tools such as out of time-order correlators (OTOCs), spectral analysis of operators and computational complexity to attack quantum many-body problems. Indeed, this article arose from our trying to answer the question: {\it What is it that makes the SYK model so special?} Is it the Majorana fermions? Or its quenched random couplings? Perhaps, it is the all-to-all $q$-fermi interactions?

An obvious starting point to answer this question would be to focus on the scrambling properties of the SYK model. Associated with the fact that the SYK${}_q$ model (for $q \geq 4$) saturates the Maldacena-Shenker-Stanford (MSS) bound \cite{Maldacena_2016} on the leading Lyapunov exponent $\lambda_{\mathrm{L}}\leq 2\pi T$ \footnote{We will work in units where $\hbar = k_B = 1$.}, it was recently argued that scrambling is better understood in terms of the growth of the size of time-evolving operators in the model \cite{Roberts:2018mnp, Qi_2019, Carrega:2020jrk}.
The idea is that in a scrambling system, the probability distribution of the size of the operator, $P_{s}(t)$, shifts towards larger operators with an initial exponential rate determined by the infinite-temperature chaos exponent.

We start instead from the seemingly very simple observation that scrambling in a many-body system is actually made up of two distinct processes: an initial small operator first grows to a sufficiently large size; at the same time, the grown operator delocalizes over the Hilbert space of large operators. Our main goal in this article will be to study the latter phase only, which we call \textit{operator delocalization}, in as simple (and universal) a setup as possible, to understand how it can be controlled. 

% \noindent
To elaborate, in this article we study the SYK$_{2}$ model. Even though this model is essentially free, the quenched random couplings $J_{ij}$ and Majorana fermions $\hat{\gamma}^{i}$ endow the system with a rich structure that has garnered much recent attention \cite{swingle_ramp_syk2, Carrega:2020jrk,Lau:2020qnl}. We go even further and define the model on a graph $G(V,E)$, consisting of a collection of vertices $V$ and edges $E\subseteq V\otimes V$ with the connectivity of the graph encoded into matrix of couplings, $J_{ij}$, now interpreted as the adjacency matrix of the graph \footnote{In the context of scrambling dynamics, the effects of the graph geometry on operator growth have been heavily studied. See, for example, \cite{Bentsen6689}}.

The key observation is that, since the SYK$_{2}$ model is free, we do not expect any operator growth through Hamiltonian evolution \cite{Carrega:2020jrk, Rabinovici:2020ryf} but this does not mean that the system is trivial. We will show that operator hopping induces nontrivial dynamics of the system which is \textit{heavily} controlled by the underlying graph. %This leads us to
We will conjecture that operator delocalization requires two ingredients, \emph{i.e.} (i) sufficiently non-local operators (either obtained by the growing dynamics of initially small operators or directly as initially large operators) and (ii) networks that are able to utilise the non-locality. 
At the technical level, we will make use of the notion of \textit{operator complexity}, introduced in \cite{Parker:2018yvk}. This Krylov, or K-complexity, $C_{K}$, describes the delocalization of an operator in a finite dimensional Hilbert space with respect to a specific basis - the Krylov basis - obtained by successive nested commutators.

The following two sections introduce and use the idea of K-complexity to provide supporting evidence for our conjecture. In section IV we show how the conjecture itself, and only the notion of operator delocalization (\textit{without} operator scrambling), can be used to understand the quantum charging advantage exhibited by the SYK quantum batteries introduced in \cite{Rossini_Rosa_PRL}.  \\

{\it II Krylov complexity for free models.---} In this section we review the notion of Krylov complexity \cite{Parker:2018yvk, Rabinovici:2020ryf, caputa2021geometry}. 
Our focus, however, will be on the differences that arise between initial operators having small or large size with the latter not usually considered when dealing with scrambling systems.

% The mathematical tool behind Krylov complexity is the recursion method and the Lanczos algorithm, adapted to the time evolution of quantum mechanical operators instead of states.
To this end, let us start with a given quantum operator, $\hat{\mathcal O}$. 
Our goal will be to efficiently describe its Hamiltonian time evolution,
\begin{equation}
    \label{eq:time_evol_operator}
    \hat{\mathcal{O}}(t) = e^{\mathrm{i} \hat{H} t} \hat{\mathcal{O}} e^{- \mathrm{i} \hat{H} t} \ .
\end{equation}
We can expand $\hat{\mathcal{O}}(t)$ over the set of the nested commutators of $\hat{\mathcal{O}}$ with $\hat{H}$, called the Krylov space, $\mathscr{K}_{\hat{\mathcal{O}}}$.
% \begin{equation}
%     \label{eq:krylov_space_def}
%     \mathcal{H}_{\hat{\mathcal{O}}} = \mathrm{span} \left\{\hat{\mathcal{O}} , \, \left[\hat{\mathcal{H}}, \, \hat{\mathcal{O}}\right] , \,  \left[\hat{\mathcal{H}}, \,\left[\hat{\mathcal{H}}, \, \hat{\mathcal{O}}\right]\right], 
%     , \, \dots\right\} \ .
% \end{equation}
Out of the Krylov set, we want to find a set of orthonormal operators \footnote{The orthornomality condition is imposed by endowing the space of the operators with the standard Frobenius product $(A|B) = \frac{1}{D} \mathrm{Tr}[A^\dagger B]$, with $D$ being the Hilbert space dimension} which can fully reconstruct $\hat{\mathcal{O}}(t)$ at any time $t$.
This subset forms the $K$-dimensional Krylov basis, $\{\hat{\mathcal{O}}_n \}_{n = 0}^{K - 1}$, where the subscript $n$ represents the number of commuting operations. 
% Out of the full Krylov set, we are interested in finding a restricted set of operators which are enough to fully reconstruct the evolved operator $\hat{\mathcal{O}}(t)$ at any time $t$.
% More in details, one endows the space of the operators with the standard Frobenius product $(A|B) = \frac{1}{D} \mathrm{Tr}[A^\dagger B]$, with $D$ being the Hilbert space dimension, and then finds a set of orthonormal operators which fully reconstruct all the Krylov space.
% Such an orthonormal set is usually called Krylov basis and denoted by $\left\{\hat{\mathcal{O}}_n \right\}_{n = 0}^{K - 1}$ where $K$ is the dimension of the Krylov basis.
In addition, the orthogonalizing coefficients, $\{b_0 \equiv 0, \, b_n \}_{n = 1}^{K - 1}$ form the set of so-called Lanczos coefficients.
In this formalism, the time-evolved operator can be expressed as
\begin{equation}
\label{eq:operator_krylov_expansion}
    \hat{\mathcal{O}}(t) = \sum_{n = 0}^{K-1} \mathrm{i}^n \varphi_n(t) \hat{\mathcal{O}}_n \ ,
\end{equation}
where $\varphi_n(t)$ satisfy the differential equations
\begin{equation}
\label{eq:phi_diff_eqns}
    \dot{\varphi}_n(t) = b_n \varphi_{n-1}(t) - b_{n+1} \varphi_{n+1}(t) \ . 
\end{equation}
In other words, knowledge of the Lanczos coefficients $b_n$ is enough to determine the operator dynamics.
To extract the information encoded in the wavefunctions $\varphi_n(t)$ in a tractable way, the K-complexity function, $C_K (t)$, can be introduced.
It computes the expected number of nested commutators for $\hat{\mathcal{O}}(t)$, %on the Krylov basis 
\begin{equation}
    \label{eq:K-complexity_def}
    C_{K}(t) = \sum_n n |\varphi_n(t)|^2 \ ,
\end{equation}
with $\sum_n |\varphi_n|^2 = 1$ and plays a pivotal role in our study.

In \cite{Parker:2018yvk} it has been argued that, by taking a \textit{simple} initial operator, \textit{i.e.} an operator that can be written as a linear combination of single Majorana fermions, the asymptotic behavior of the coefficients $b_n$, for $n \lesssim \mathrm{log} D$ can be used to diagnose the chaotic/integrable nature of the Hamiltonian under investigation.
In particular, for chaotic models one has $b_n \propto n$ \footnote{With a logarithmic correction for models defined in $1$ dimension which has been studied numerically in very great detail in \cite{noh2021operator}.} while for integrable models in general one has $b_n \propto n^\alpha$, for some $\alpha < 1$. 
The extreme case of free models gives $b_n \sim O(1)$.

A crucial consequence of equations \eqref{eq:operator_krylov_expansion} and \eqref{eq:phi_diff_eqns} is that, for short enough times, $\hat{\mathcal{O}}(t)$ and $C_K(t)$ are mostly controlled by the $b_n$'s with small $n$.
We then conjecture that at early times there could well be integrable or free models sharing similar physical properties with chaotic models.
However, this intuition applies only when considering initial \textit{large} operators, involving products of many $\hat{\gamma}_i$. 
Such large operators are required because free models, such as SYK$_2$, do not produce operator growth dynamics; they simply translate operators in the operator space \cite{Carrega:2020jrk, Rabinovici:2020ryf}. 
However, when starting with operators of large size and not requiring any further operator growth, the resulting hopping dynamics can be very close to be chaotic at early times.
In particular, we expect that the dynamics of large operators will be highly dependent on the connectivity of the graph which defines the model. In the following section we test this intuition.\\

%\textcolor{red}{In the final version of the paper we will make the argument above stronger with a plot (perhaps in an appropriate appendix). In that plot we want to compute, by averaging the $b_n$s over the disorder couplings, the first, say, $20$ averaged $b_n$ for the SYK$_2$ model and for any given size. The idea is to compare these coefficients as a function of the size against the linear growth that we have for SYK$_4$ and use it as a further indication that the early time behavior can mimic a chaotic dynamics. Notice that the $b_n$s, if averaged over the random couplings, can be computed analytically for SYK$_2$ at large $L$.}

%%%%%%%%%%%%%%%%%%%%%%%%%%%%%%%%%%%%%%%%%%%%%%%%%%%%%%%%%%%%%%%%%%%%%%
%\section{Operator delocalization without scrambling}
%\label{sec:delocalization_properties}
%%%%%%%%%%%%%%%%%%%%%%%%%%%%%%%%%%%%%%%%%%%%%%%%%%%%%%%%%%%%%%%%%%%%%%

{\it III Operator delocalization without scrambling.---} We now study the time evolution of the K-complexity function, $C_K (t)$, for particular deformations of the SYK$_2$ model, controlled by the topologies of the graphs over which the models live.
Our goal here is to study the extent to which $C_K (t)$ is a good probe to distinguish the properties of the graphs governing the dynamics of the model under investigation.

%%%%%%%%%%%%%%%%%%%%%%%%%%%%%%%%%%%%%%%%%%%%%%%%%%%%%%%%%%%%%%%%%%%%%%
%\subsection{Models}
%\label{ssubec:models}
%%%%%%%%%%%%%%%%%%%%%%%%%%%%%%%%%%%%%%%%%%%%%%%%%%%%%%%%%%%%%%%%%%%%%%

Consider a set of SYK$_2$ models defined on graphs, beyond the standard choice of complete graphs that characterize the all-to-all interactions typical of SYK physics. The SYK model is a quantum mechanical model of Majorana fermions in $1$ dimension, consisting of operators $\hat{\gamma}^i$, $i = 1 ,\, 2,\, \dots ,\, L$, satisfying the Clifford algebra $\left\{ \hat{\gamma}^i , \, \hat{\gamma}^j \right\} = \delta^{ij}$, with random $q$-body interactions. In this article, we focus on the case of a quadratic Hamiltonian,
\begin{equation}
    \label{eq:SYK_hamiltonian}
    \hat{H}^{(2)} = \mathrm{i} \sum_{i < j} J_{ij} \hat{\gamma}^{i} \hat{\gamma}^{j} \ .
\end{equation}
In the usual formulation of the model, the all-to-all coupling constants $J_{ij}$ are randomly extracted from a Gaussian distribution, with vanishing mean and variance $\langle J_{ij}^2 \rangle = J^2/L$, and where the constant $J^2$ has the dimension of energy. In what follows, we set $J=1$.

Similar to the sparse SYK models of \cite{xu2020sparse, PhysRevD.103.106002, caceres_sparse_2021}, we consider models living on graphs different from the complete graph case described above.
To implement this, we replace the matrix of couplings $J_{ij}$ with the adjacency matrix, $A_{ij}$, of a given graph, $G(L, E)$, with $L$ vertices (one for each Majorana fermion) and $E$ edges. Finally, we multiply each non-vanishing entry of $A_{ij}$ (with $i < j$) with a random number extracted from a Gaussian distribution having vanishing mean and variance $\frac{L - 1}{2 n_E}$, with $n_E$ denoting the number of edges in $G(L, E)$.
This procedure produces a new matrix of couplings, $\tilde J_{ij}$, which defines a quadratic Hamiltonian, 
\begin{equation}
    \label{eq:SYK_hamiltonian_graph}
    \hat{H}^{(2)}_{G(L, E)} = \mathrm{i} \sum_{i < j} \tilde J_{ij} \hat{\gamma}^{i} \hat{\gamma}^{j} \ ,
\end{equation}
in the same fashion as for the complete graph, equation \eqref{eq:SYK_hamiltonian}. Among these networks, {\it small-world} graphs \cite{watts_collective_1998} form a distinguished subset. Produced by the so-called Watts-Strogatz algorithm, they parametrically interpolate between a regular lattice and a random Erd\"os-Renyi graph. The algorithm is specified in terms of two numbers: an integer $k$ and a probability value $p\in[0,1]$. For a given value of $k$, it starts with a regular circulant lattice in which each vertex is connected to its $2k$ nearest neighbours \footnote{In the language of graph theory, $2k$ is the fixed degree of each node in the graph.}. 
Edges are subsequently rewired at random with probability $p$, avoiding self-loops, edge duplication and keeping the graph connected.
When $p = 0$ and $k = 1$, the resulting SYK$_2$ model is equivalent to a nearest neighbors tight-binding system, from which no interesting dynamics can arise.
On the other hand, by dialling the value of $p$, networks become highly interconnected and the mean distance between two edges can be very short.
We will show that this geometry change has huge impact on the SYK$_2$ physics.

\begin{figure}[t!]
\begin{center}
    \includegraphics[width=1\hsize]{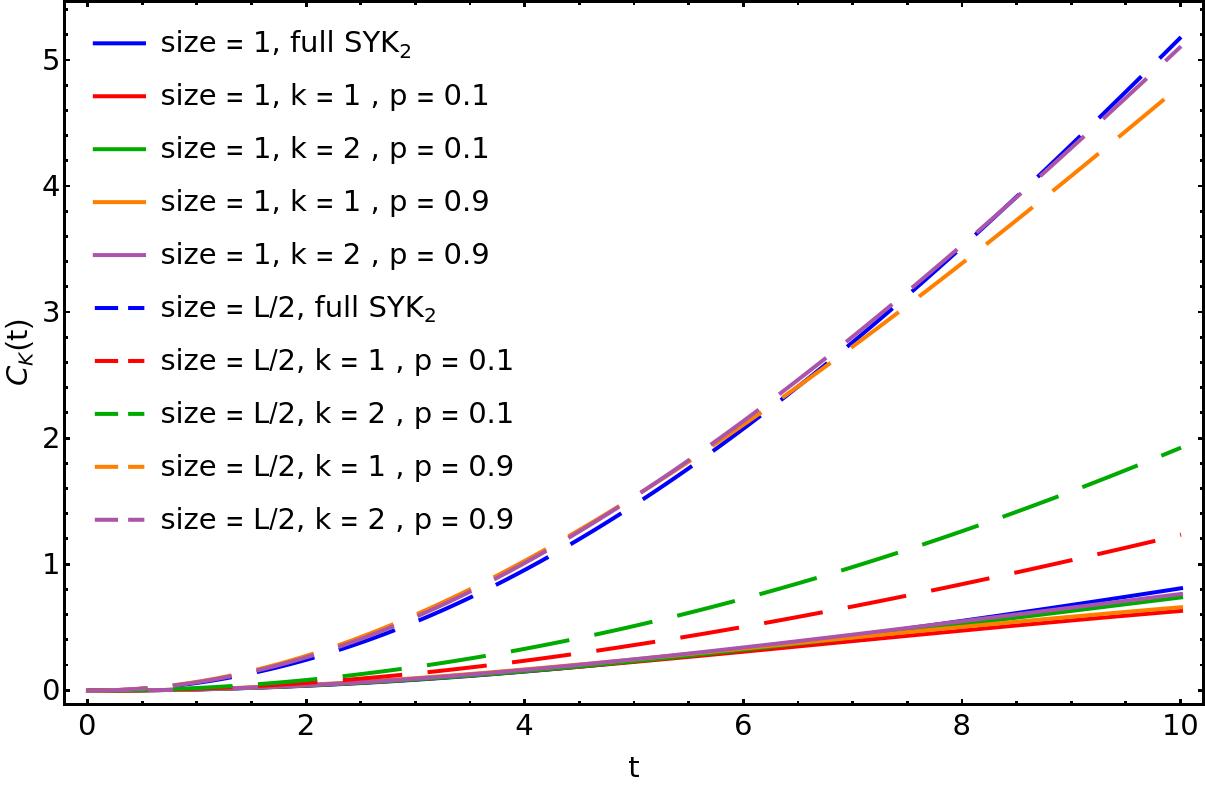}\\
    \caption{$C_K(t)$ for systems having $L = 24$. $C_K(t)$ is computed for small (size $1$) and large (size $L/2$) operators, for the full SYK$_2$ model, compared against the Watts-Strogatz Hamiltonians having $k = 1, \, 2$ and both \textit{low} and \textit{large} rewiring probability ($p = 0.1$ and $p = 0.9$, respectively). The results are averaged over $1000$ realizations of disorder and graph.}
\label{fig:complexity_time_evolution}
\end{center}
\end{figure}

Given these preliminaries, we have computed the early-time evolution of $C_k(t)$, \textit{i.e.} for times much shorter than the saturation time, for several choices of $k$ and $p$ and for both small operators (operators of size $1$, \textit{i.e.} the simple operator $\hat \gamma^1$) and for large operators (operators having \textit{extensive} size equal to $L/2$, \textit{i.e.}  $\hat{\mathcal O}^{(L)} \equiv \prod_{i = 1}^{L/2} \hat \gamma^i$). For comparison, we have also computed the time evolution of the $K$-complexity for the fully connected SYK$_2$ model. In all cases, to remove a possible source of spurious effects, we have normalized the Hamiltonians to have unit bandwidth, \textit{i.e.} we imposed that the difference between the largest and the smallest eigenvalue is equal to $1$. 
The results are reported in figure \ref{fig:complexity_time_evolution}. 

There are a number of points we wish to draw attention to. First, the underlying graph plays essentially no role for {\it small} operators. 
In all cases, $C_{K}(t)$ displays a slow growth and the full SYK curve is just barely distinguishable from the small-world curves, at both high and low rewiring probabilities.
This is just another manifestation of the fact that, irrespective of the underlying quantum network, the SYK$_2$ model is not a scrambling system. As such, it cannot create any operator size and so, does not utilize the connectivity of the underlying graph. 
The situation changes drastically when \textit{large} operators are involved: here, we see that the evolution of the K-complexity function which, as already mentioned, quantifies the delocalization properties of the system, varies dramatically with a change of the graph topology. In particular, for low re-wiring probabilities, the Watts-Strogatz Hamiltonians exhibit much smaller values of $C_K(t)$ compared to the full SYK$_2$ Hamiltonian. On the other hand, when $p$ is large, the Krylov-complexity for quantum small-world graphs is essentially  equivalent to the corresponding function for the full SYK$_2$ model.

\begin{figure}[h!]
\begin{center}
    \includegraphics[width=1\hsize]{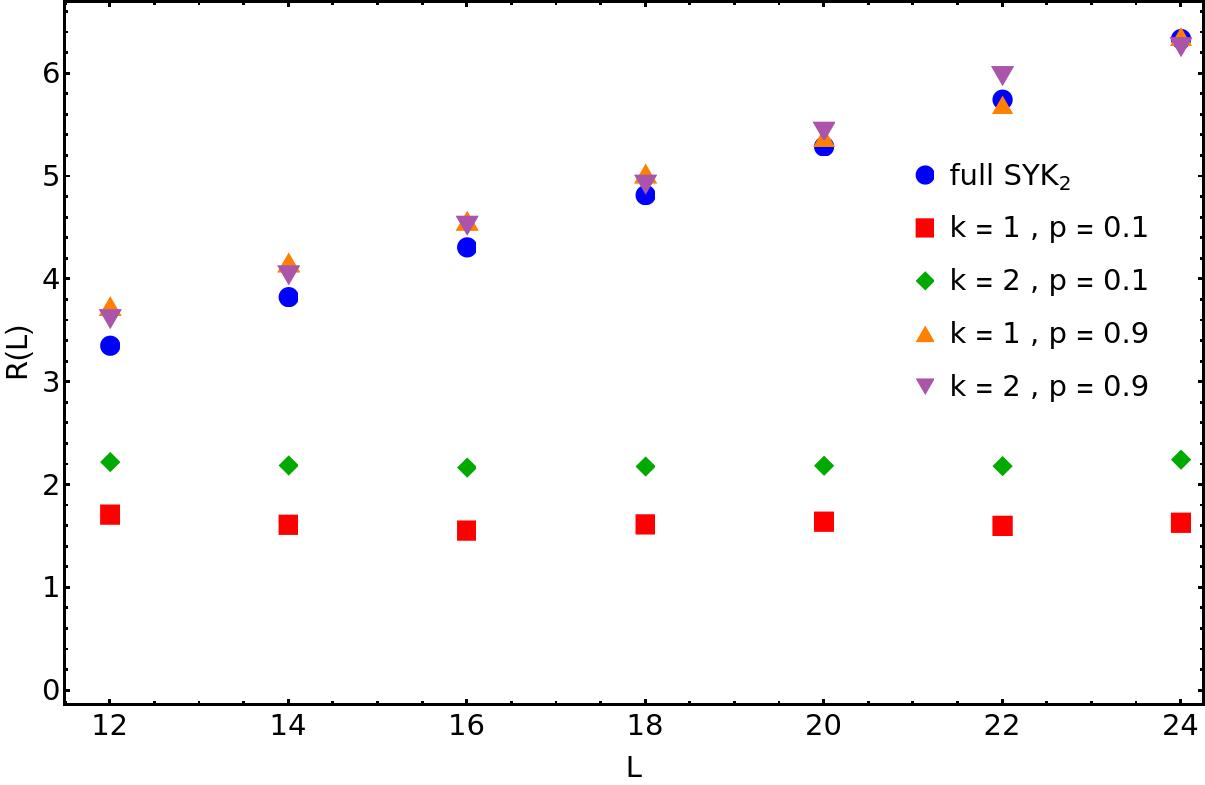}\\
    \caption{The quantity $R(L)$, computed at different system sizes and for the full SYK$_2$ model, compared against the Watts-Strogatz Hamiltonians having $k = 1, \, 2$ and both \textit{low} and \textit{large} rewiring probability ($p = 0.1$ and $p = 0.9$, respectively). All the results are averaged over $1000$ different realizations of disorder and underlying graph.}
\label{fig:complexity_ratios}
\end{center}
\end{figure}

Another interesting figure of merit to quantify the ability of a given graph to delocalize large operators is given by the ratio $R(t)$, between $C_K(t)$ for operators of size $L/2$ and operators of size $1$. As is evident from its definition, $R(t)$ measures how good a given graph is in utilizing operators of large size, normalized by the delocalizing properties computed for small operators.
Interestingly, we note that $R(t)$ is essentially \textit{time-independent}. It therefore makes sense to consider instead the quantity $R(L)$, defined as such a constant ratio and computed as a function of the system size, $L$.
Our results are depicted in figure \ref{fig:complexity_ratios}.
The main feature of note is that the difference between the highly connected and poorly connected graph is now \textit{quantitatively} clear; for highly connected graphs, $R(L)$ scales with the system size, a feature shared with the full SYK$_2$ Hamiltonian. 
On the other hand, poorly connected graphs do not show any scaling behavior for $R(L)$. 
This lacking of a scaling happens because, without long range interactions, the early-time physics is dominated by the \textit{local} features of the graph and, in particular, the system size does not affect the dynamics.

Taken together, these results show that, when scrambling dynamics is absent, the topology (and in particular the connectivity) of the graph over which the model is defined becomes the crucial ingredient to understand how large operators delocalize under quantum evolution.
We note also that the total number of connections in the model is irrelevant. This is seen from the similarities in delocalization properties of the Watts-Strogatz Hamiltonians and the full SYK$_2$ model at large $p$, where the latter has $O(L^2)$ edges against the $O(L)$ edges of the former.
This property is the analog, for non scrambling models, of the results discussed in \cite{xu2020sparse, PhysRevD.103.106002, caceres_sparse_2021} for the sparse SYK$_4$ models, which exhibit similar behavior to the full model, but with significantly fewer non-vanishing couplings. 

%%%%%%%%%%%%%%%%%%%%%%%%%%%%%%%%%%%%%%%%%%%%%%%%%%%%%%%%%%%%%%%%%%%%%%
% \section{Application: the quantum charging advantage of SYK-like quantum batteries}
% \label{sec:charging power_SYK}
%%%%%%%%%%%%%%%%%%%%%%%%%%%%%%%%%%%%%%%%%%%%%%%%%%%%%%%%%%%%%%%%%%%%%%
{\it IV An application to the quantum charging advantage of SYK-like quantum batteries.---} The K-complexity framework can be used to understand the quantum charging advantage of SYK-like quantum batteries \cite{Rossini_Rosa_PRL}. We report here the main results, and defer details to the Supplemental Material. 

An SYK quantum battery (see \cite{sai_review, dutta_review} for an overview of the topic) is built by considering a system prepared in the ground state, $\ket{0}$,  of a static initial Hamiltonian of the form $\hat{H}_0 = h \sum_{i = 1}^{L/2} \hat \sigma_{i}^x$, where $h$ denotes a constant magnetic field, oriented along the $x$-axis (which we will set equal to $1$), and $\hat \sigma_{i}^a$, with $a = x,\,y,\, z$, are the usual Pauli operators, defined on a spin chain of length $L/2$. At $t = 0$ the system is suddenly coupled, via a standard Jordan-Wigner map, to an SYK Hamiltonian and evolved under the quantum quench. The average charging power of the battery reads
\begin{align}
    \label{eq:average_power_def}
    % P_{\mathrm{av}} (t) & \equiv \frac{1}{t} (\overline{E(t)} - E_g)  \ \nonumber \\
    % &= \frac 1t \left( \overline{\bra{\psi(t)}  \hat{H}_0  \ket{\psi(t)}} - \bra{0}  \hat{H}_0  \ket{0} \right) \ ,
    P_{\mathrm{av}} (t) =  \frac{\overline{\bra{\psi(t)}  \hat{H}_0  \ket{\psi(t)}} - \bra{0}  \hat{H}_0  \ket{0}}{t} \ ,
\end{align}
where $\ket{\psi(t)}$ denotes the evolved state at time $t$, $\bra{0}  \hat{H}_0  \ket{0}$ is the ground state energy and $\overline{\bra{\psi(t)}  \hat{H}_0  \ket{\psi(t)}}$ measures the energy stored in the battery (averaged over disorder and possible graph realizations).
By quenching with an SYK$_4$ Hamiltonian (rescaled to unit bandwidth), it was found in \cite{Rossini_Rosa_PRL} that the maximum value of the average power, $P_{\mathrm{max}}$, scales with $L$, signalling a quantum charging advantage \cite{Binder_2015, campaioli_goold_advantage, KRS_advantage}.

A crucial point, first noticed in \cite{Carrega:2020jrk} and proven in full generality in \cite{KRS_advantage}, is that such charging advantage strongly relies on the fact that the $\hat \sigma^x_i$, when written in terms of $\hat{\gamma}^i$, have \textit{very large} size. This in turn suggests that charging advantage may also be obtained from SYK$_2$ models on highly connected graphs.

This intuition can indeed be confirmed for the full SYK$_2$ model: after considering the ensemble average over the Gaussian couplings, $P_{\mathrm{av}}(t)$ takes the form
\begin{equation}
    \label{eq:P_av_complexity}
    P_{\mathrm{av}} (t) \propto  \frac{\overline{\varphi_0^{(\hat{H}_0)}(t)} - 1}{t} \ ,
\end{equation}
where $\overline{\varphi_0^{(\hat{H}_0)}(t)}$ is the (averaged over disorder) first wavefunction, defined as in equation \eqref{eq:operator_krylov_expansion}, in the Krylov expansion of  $\hat{H}_0$. In particular, equation \eqref{eq:P_av_complexity} expresses the relationship between the average power and the delocalizing properties of the quench Hamiltonian, quantified by the Krylov complexity.
\begin{figure}[t!]
\begin{center}
    \includegraphics[width=1\hsize]{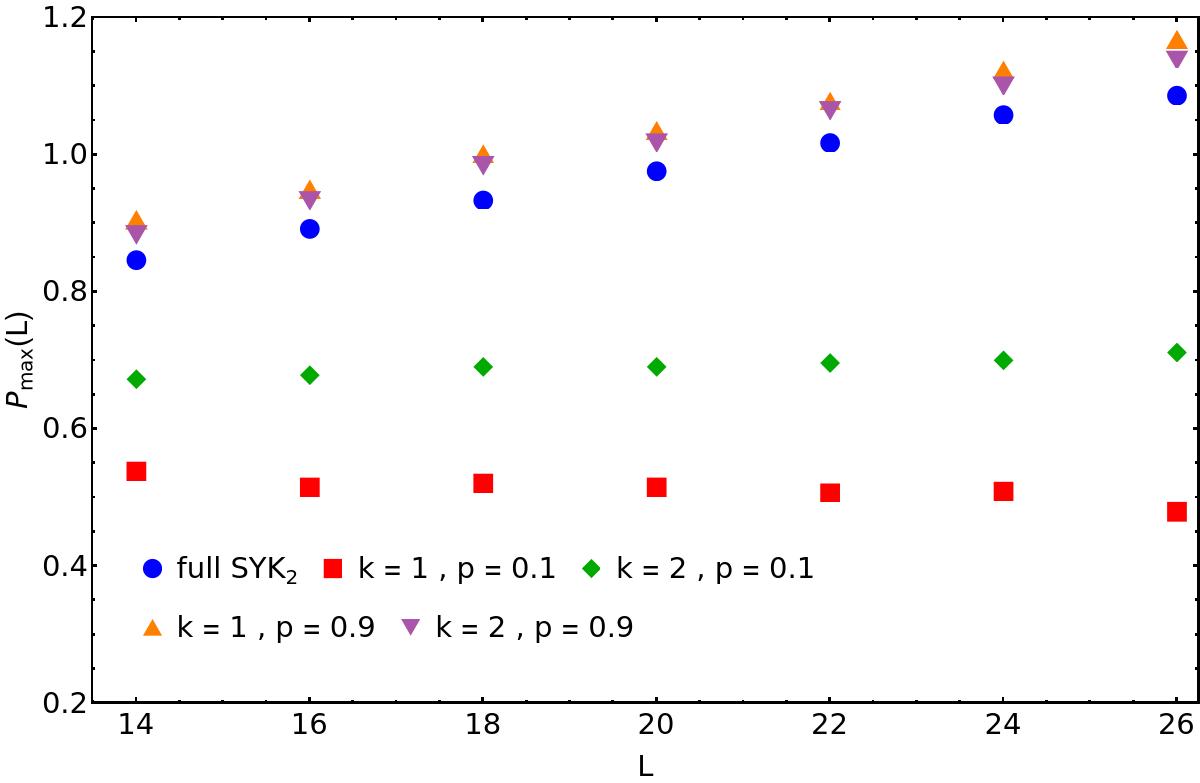}\\
    \caption{The maximum charging power, $P_{\mathrm{max}}(L)$ for the same models considered in figure \ref{fig:complexity_ratios}.}
\label{fig:maxPower}
\end{center}
\end{figure}
For the more general Watts-Strogatz Hamiltonians equation \eqref{eq:P_av_complexity}, which is valid for a \textit{fixed} graph topology, must be supplemented with an additional average over the graph topology. As an illustration, figure \ref{fig:maxPower} displays the result of our numerical computation of $P_{\mathrm{max}}(L)$ for several Watts-Strogatz Hamiltonians, compared against the full SYK$_2$ charging power. This clearly matches the analogous results obtained in figure \ref{fig:complexity_ratios} and demonstrates that K-complexity and the delocalization properties are indeed the relevant quantities to understand the quantum charging advantage of SYK quantum batteries. The scaling, exhibited by highly connected graphs, shows that interaction connectivity is sufficient to obtain such an advantage.

{\it V Conclusions.---}SYK-like models defined on graphs offer a versatile and novel class of quantum systems to explore many-body localization, thermalization and chaos \cite{xu2020sparse, PhysRevD.103.106002, caceres_sparse_2021, PhysRevB.99.054202}.
Previous studies that investigated  the preservation of chaotic properties by the graph topology, have focused primarily on the strongly interacting SYK$_4$ Hamiltonians.
In this paper, to better disambiguate spectral properties from those hinging on the (hyper)graph structure, we have focused instead on the free SYK$_2$ model defined on various graphs.
The spectral properties of the system are then rather trivial, and any non-triviality must be a consequence of the underlying graph structure.
Here, we have shown that, as long as operators of \textit{sufficiently large size} are taken into account, the dynamics is far from trivial and in particular the K-complexity function, $C_K(t)$, is highly dependent on the geometry and  connectivity of the graph.
In turn, this observation has led us to propose the notion of ``operator delocalization'', describing how large operators delocalize under the operator hopping dynamics \cite{Carrega:2020jrk, Rabinovici:2020ryf}.

As an application of these ideas, we have shown also that the quantum charging advantage of SYK quantum batteries, found in \cite{Rossini_Rosa_PRL}, is a direct consequence of operator delocalization and, as such, relies only on the underline graph topology.

There are many intriguing future directions worth pursuing.  It would be of significant interest to find other physical quantities which are highly sensitive to operator delocalization only. Another interesting direction would be to understand how operator delocalization depends on the statistics of the evolving operators. Of particular interest, given their relevance in condensed matter systems, would be the study of SYK-like models build from parafermionic operators \cite{alicea_review_parafermions, PhysRevA.89.012328, rossini_carrega_parafermions}.

{\it Acknowledgements.---}
We thank Matteo Carrega for collaboration during the early stages of this project.
DR would like to thank Juyeon Kim, Dominik \v{S}afr\'{a}nek and Ruth Shir for collaboration on related projects. 
JM thanks the organisers of Strings 2021 for the opportunity to preview our results to a stimulating audience.
JO acknowledges the support by the grants FPA2017-88915-P and SEV-2016-0588 from MINECO and 2017-SGR-1069 from DURSI. IFAE is partially funded by the CERCA program of the Generalitat de Catalunya.
JK acknowledges the support by the NSF grant PHY-1911298 and the Sivian fund.
DR acknowledges the support by the Institute for Basic Science in Korea (IBS-R024-Y2 and IBS-R024-D1). JM is partly funded by the National Research Foundation of South Africa.
\bibliography{main}

\clearpage

\setcounter{section}{0}
\setcounter{equation}{0}%
\setcounter{figure}{0}%
\setcounter{table}{0}%

\setcounter{page}{1}

\renewcommand{\thetable}{S\arabic{table}}
\renewcommand{\theequation}{S\arabic{equation}}
\renewcommand{\thefigure}{S\arabic{figure}}
\onecolumngrid

\begin{appendices}

In this Supplemental Material we elaborate on some details of the quantum charging advantage of  SYK-like quantum batteries, including a perturbative analytical proof of the existence of the charging advantage in the thermodynamic limit. Moreover, we present a specific example of Small-World SYK Hamiltonian with star graph topology, which can be easily solved analytically and provides further confirmation of the results reported in the main text.

\section{More on SYK-like quantum batteries}

Here we present more details on the properties of the SYK quantum batteries, focusing on the case of the full SYK$_2$ Hamiltonian. However, when discussing the extensive charging advantage in the thermodynamic limit we will show how the argument can be extended to include the case of the SYK$_4$ quantum batteries studied in \cite{Rossini_Rosa_PRL}.

%%%%%%%%%%%%%%%%%%%%%%%%%%%%%%%%%%%%%%%%%%%%%%%%%%%%%%%%%%%%%%%%%%%%%%%%%%%%%%%%%%%
\subsection{The charging protocol}
\label{subsec:charging_protocol}
%%%%%%%%%%%%%%%%%%%%%%%%%%%%%%%%%%%%%%%%%%%%%%%%%%%%%%%%%%%%%%%%%%%%%%%%%%%%%%%%%%%

We start by recalling the charging protocol used to inject energy into the quantum batteries under investigation.
Toward this end, consider a quantum battery made of $L/2$ cells (essentially, just $L/2$ qubits). We consider the \textit{discharged} battery as given by a static, non-interacting, Hamiltonian together with its ground state
\begin{equation}
    \label{eq:discharged_battery}
    \hat{H}_0 = h \, \sum_{i = 1}^{L/2} \hat \sigma_i^a \  \Longrightarrow \ket{0} \ , \ \textrm{with} \ \hat{H}_0 \ket{0} = E_0 \ket{0} \ ,
\end{equation}
where $h$ denotes the strength of a constant magnetic field (which we set to be equal to $1$), whose orientation is along the axis $a = x, \, y, \, z$. The operators $\hat \sigma_i^a$ are the usual Pauli operators at the site $i$ and $E_0$ denotes the ground state energy, $\bra{0}\hat H_0 \ket{0}$.
In particular, we will focus on the cases in which the external magnetic field is aligned along the $x$ or $z$ axes and denote the corresponding batteries as ``$x$-battery'' and ``$z$-battery'', respectively.

To charge the batteries, we use a double sudden quench. 
By this, we mean that at $t=0$ we instantaneously change the Hamiltonian to $\hat{H}_1$, which does not have $\ket{0}$ as an eigenstate, to induce  non-trivial dynamics. 
After a certain amount of time, we switch back to the original Hamiltonian $\hat{H}_0$.
% The charging protocol is therefore controlled by the following, time-dependent,  Hamiltonian
% \begin{equation}
%     \label{eq:charging_protocol_def}
%     \hat{H}(t) \equiv \hat{H}_0 + \lambda(t) \left(\hat{H}_1 - \hat{H}_0 \right) \ ,
% \end{equation}
% where the function $\lambda(t)$ is equal to $1$ when $0 \geq t \geq \tau$, and $0$ otherwise.

As a consequence of the non-trivial dynamics induced by the quench Hamiltonian $\hat{H}_1$, the initial state $\ket{0}$ has evolved into a superposition of eigenstates of $\hat{H}_0$, generically denoted by $\ket{\psi(t)}$.
Hence, the mean value of the energy stored in the battery, as measured by
\begin{equation}
    \label{eq:energy_stored_def}
    E(t) \equiv \bra{\psi(t)} \hat{H}_0 \ket{\psi(t)} - E_0 \ ,
\end{equation}
is positive.
% Let us note that, since for $t > \tau$ the time evolution is again controlled by $\hat{H}_0$, the energy stored in the battery remains constant for $t > \tau$ (although the quantum state has a non-trivial time evolution).
Our goal is to find the conditions under which the optimal charging power, $P_{\mathrm{max}}$, defined by the equation
\begin{equation}
    \label{eq:optimal_power_def}
    P_{\mathrm{max}} \equiv \max_{t} P_{\mathrm{av}}(t) =  \max_{t} \left(\frac{E(t)}{t} \right) \ ,
\end{equation}
shows a scaling with the system size $L$ which cannot be reached by any possible classical device, \textit{i.e.} reflecting an actual \textit{extensive quantum charging advantage}. By rescaling the quench Hamiltonian as
\begin{equation}
    \label{eq:rescaling_quench}
    \hat{H}_1 \to \frac{\hat{H}_1}{\mu(\hat H_1)} \ ,
\end{equation}
where $\mu(\hat H_1)$ denotes the bandwidth of $\hat{H}_1$, \textit{i.e.} the difference between its maximum and minimum energies, it was numerically shown in \cite{Rossini_Rosa_PRL} that a  quantum battery in which $\hat{H}_0$ is given by the $x$-model and the quench Hamiltonian $\hat{H}_1$ is given by an SYK$_4$ Hamiltonian shows a non-trivial, positive, scaling and thus displays an extensive quantum charging advantage. As shown in the main text, such an advantage \textit{does not} require the full chaotic SYK$_4$ model and can, in fact, be achieved with the simpler quadratic model also.

\subsection{Details on the couplings: the Jordan-Wigner map and locality in Majorana space}
\label{subsec:JW_locality}

We now turn to the details of the Jordan-Wigner (JW) map that we have used throughout the paper, and that determines how local (or non-local) the spin operators are in Majorana space.
In practice, we need to express the Majorana variables, appearing in the definition of the SYK model, in terms of the spin variables (or vice-versa) describing the static battery, $\hat{H}_0$, equation \eqref{eq:discharged_battery}.
This procedure encodes the fermionic variables into the qubits and effectively represents a degrees of freedom in building the particular models of quantum batteries under considerations.
For our purposes we will adopt the JW map
\begin{align}
    \label{eq:JW_map_def}
    & \hat{ \gamma}^{2j - 1} = \frac{1}{\sqrt 2} \left( \prod_{i = 1}^{j - 1} \hat{\sigma}_i^z \right) \, \hat{\sigma}_j^x \ , \nonumber \\
    & \hat{ \gamma}^{2j} = \frac{1}{\sqrt 2} \left( \prod_{i = 1}^{j - 1} \hat{\sigma}_i^z \right) \, \hat{\sigma}_j^y \ .
\end{align}

\begin{figure}[th!]
\begin{center}
    \includegraphics[width=.5\hsize]{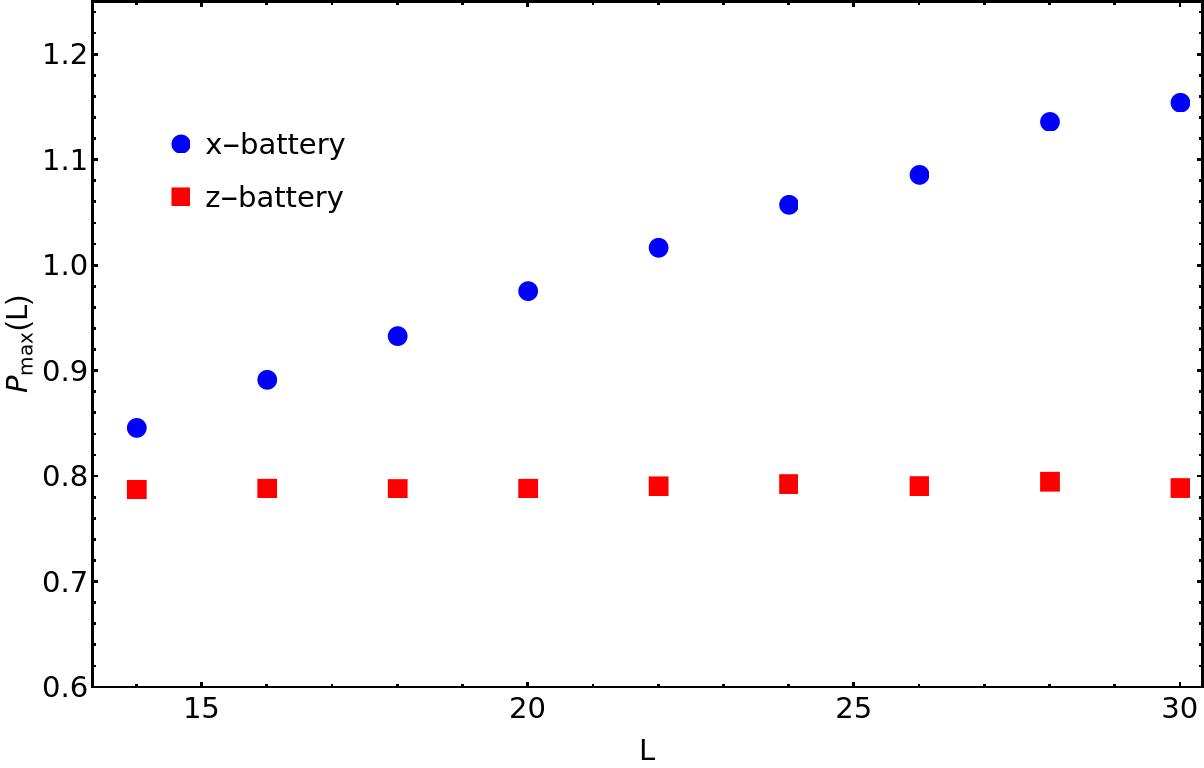}\\
    \caption{The maximum charging power, $P_{\mathrm{max}}(L)$ for the $x$-battery and for the $z$-battery, both based on a quench Hamiltonian given by the full SYK$_2$ model.}
\label{fig:maxPowerXvsZ}
\end{center}
\end{figure}

Now, the crucial observation, described in detail in \cite{Carrega:2020jrk}, is that the \textit{operator size} of the static Hamiltonians, $\hat{H}_0^a$, once rewritten in the Majorana fields by means of equation \eqref{eq:JW_map_def}, is \textit{highly dependent} on whether the index $a$ is identified with $x$ or $z$.
To see why this is true, we notice that the simple operators $\hat{\sigma}_i^{a}$ which constitute the static Hamiltonians, are no longer fundamental operators when rewritten in terms of the Majorana variables, $\hat{\gamma}^i$.
In particular, their sizes differ significantly when passing from $a = z$ to $a = z$ and we have
\begin{equation}
    \label{eq:size_pauli_operators}
    \hat{\sigma}_i^z = -2 \mathrm{i} \, \hat{\gamma}^{2 i - 1} \hat{\gamma}^{2i} \ , \qquad \hat{\sigma}_i^x = \left(\sqrt{2}\right)^{2i - 1} (- \mathrm{i})^{i - 1} \prod_{p = 1}^{2 i - 1} \hat{\gamma}^p \ ,
\end{equation}
which shows that $\hat{\sigma}_i^z$ has size $2$ when expressed in terms of the Majorana variables, while $\hat{\sigma}_i^x$ has size $2 i - 1$ when expressed in terms of the Majorana variables.
This large difference in system size is  reflected in the maximal charging power, in agreement with the general theorem found in \cite{KRS_advantage} and as we numerically show in figure~\ref{fig:maxPowerXvsZ}.
As can clearly be seen, the maximum charging power exhibits an \textit{extensive} scaling only for the $x$-battery, confirming that the operator size is crucial in order to obtain such an advantage.

We now turn to show that the extensive charging advantage, and its deep relation with operator size, can be established in the thermodynamic limit via a perturbative computation. This in turn can be extended to include the case of SYK$_4$ quantum batteries as well.

\subsection{The quantum charging advantage in the thermodynamic limit}

The average charging power $P_{\mathrm{av}}(t)$, as expressed in equation \eqref{eq:average_power_def}, once interpreted in the Heisenberg picture can be perturbatively evaluated by means of a BCH-like formula
\begin{equation}
    \label{eq:average_power_heisenberg}
    P_{\mathrm{av}}(t) = \frac 1t \left( \bra{0} \overline{ \exp(\mathrm{i} \frac{\hat{H}_1}{\mu(\hat H_1)} t) \hat{H}_0 \exp(-\mathrm{i} \frac{\hat{H}_1}{\mu(\hat H_1)} t)} \ket{0} - \bra{0} \hat{H}_0 \ket{0} \right) \ ,
\end{equation}
where, for the moment, $\frac{\hat{H}_1}{\mu(\hat H_1)}$ is assumed to be the full (rescaled) SYK$_2$ Hamiltonian, \textit{i.e.} the quadratic SYK model living on the complete graph with $L$ vertices and, as usual, we denote with an overbar the average over the Gaussian couplings.

As a result of this averaging over the Gaussian couplings, equation \eqref{eq:average_power_heisenberg} simplifies dramatically, since all the Gaussian couplings entering on its right hand side must be coupled in pairs or they vanish \textit{on average}, a fact explored at length in \cite{Carrega:2020jrk}.
Taking this simplification into account we end up with the following formula involving odd power of $t$ only
\begin{align}
    \label{eq:P_inst_gen_massaged}
    & P_{\mathrm{av}}(t) =  - \frac{ t}{2 \, \mu^2(\hat H_1)} \bra{0}\overline{\left[\hat{H}_1 , \left[ \hat{H}_1 ,  \hat{H}_0 \right] \right] } \ket{0} + \frac{t^3}{4! \, \mu^4(\hat H_1)} \bra{0} \overline{\left[\hat{H}_1 ,\left[\hat{H}_1 ,\left[\hat{H}_1 , \left[ \hat{H}_1 ,  \hat{H}_0 \right] \right]\right] \right] } \ket{0} + O(t^5) \ .
\end{align}

This in turn can be used to compute, order by order, the average charging power. To be concrete, we will focus on the more cumbersome $x$-battery with the results easily extended to the simpler $z$-battery case which exhibits no extensive quantum advantage. In the case of the $x$-battery however,the static Hamiltonian, $\hat{H}_0$, takes the  form
\begin{equation}
    \label{eq:H0_xmodels}
    \hat{H}_0 \equiv \sum_{k = 1}^{L/2} \hat{\sigma}_k^x = \sum_{k = 1}^{L/2} \hat{\mathcal{O}}^{(2 k - 1)} \ ,
\end{equation}
where we have made use of equation \eqref{eq:size_pauli_operators} to recall that the static Hamiltonian, in the Majorana basis, can be expressed as a sum of operators having all the odd sizes from $1$ to $L - 1$.
As described in \cite{Carrega:2020jrk}, the expression \eqref{eq:H0_xmodels}, once plugged into the nested commutators \eqref{eq:P_inst_gen_massaged}, and taking into account the average over the Gaussian couplings, gives rise to a \textit{return amplitude}. Loosely speaking, this computes the likelihood that the evolved operator returns to its original form.
In details, we have computed equation \eqref{eq:P_inst_gen_massaged} up to $8$ nested commutators.
Putting together these pieces, gives the following expression for the average power
\begin{align}
    \label{eq:P_inst_SYK2_pert}
    & P_{\mathrm{av}} (t) = \frac{1}{ 2 L \,  \mu(\hat{H}_1)^2} \left( \sum _{k=1}^{L/2} (2 k-1) (-2 k+ L+1) \right) \, t + \nonumber \\
    & - \frac{1}{4! L^2 \, \mu(\hat{H}_1)^4} \left( \sum _{k=1}^{L/2} (2 k-1) (-2 k+ L+1) \left((6 k-4) L-3 (2 k-1)^2+2\right) \right) t^3 + \nonumber \\
    & + \frac{1}{ 6! L^3 \, \mu(\hat{H}_1)^6 }  \left(\sum _{k=1}^{L/2} 5 (2 k-1) (L-(2 k-1)) \left(-6 (2 k-1) (\frac L2-1) ( L-(2 k-1))+ \right. \right. \nonumber  \\
    & + \left. 3 (2 k-1)^2 ( L-(2 k-1))^2+4 (\frac L2 -1)^2\right) \Biggr) t^5 + \nonumber \\
    & - \frac{1}{ 8! L^4 \,   \mu(\hat{H}_1)^8 } \left(\sum _{k=1}^{L/2} 7 (2 k-1) (-2 k+L+1) \Bigl(-4 (2 k-1) \Bigl(\frac{25}{4} L^2-\frac{51}{2} L+32\Bigr) (2 k- L-1)+ \Bigr. \right. \nonumber \\
    &  -15 (2 k-1)^3 (-2 k+ L+1)^3-60 (1-2 k)^2 (\frac L2-1) (-2 k+ L+1)^2-8 L^3+49 L^2 - 118 L+80 \Bigr) \Biggr) t^7 + O(t^9)\ .
\end{align}

\begin{figure}[th!]
\begin{center}
    \includegraphics[width=.5\hsize]{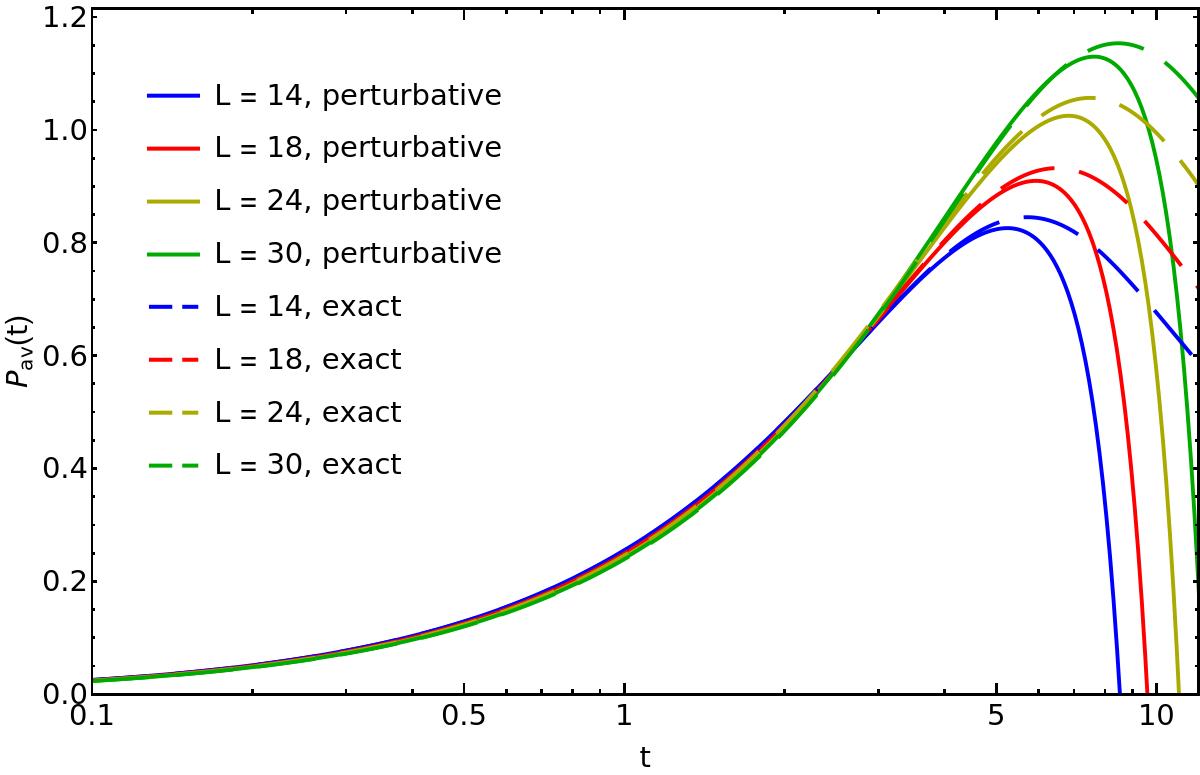}\\
    \caption{The comparison between the average power, $P_{\mathrm{av}}(t)$ as computed via the perturbative formula, equation \eqref{eq:P_inst_SYK2_pert} (solid lines) and via exact diagonalization (dashed lines).}
\label{fig:power_plot _SYK2 _xmodel _pertvsnum}
\end{center}
\end{figure}

Let us first check how \eqref{eq:P_inst_SYK2_pert} compares with the average numerical power curves.
To this end, we have to estimate the Hamiltonian bandwidth, $\mu(\mathcal H_1)$, as a function of $L$.
We get this estimate by means of a linear fit of the numerical bandwidth at finite $L$.
Using this estimation for $\mu(\hat{H}_1)$, we report the comparison between the numerical curves and the perturbative approximation for the average power in figure~\ref{fig:power_plot _SYK2 _xmodel _pertvsnum}.

Notice that, up to the time of maximal power, the agreement is extremely good and, more importantly, the extensive scaling is clearly reproduced. This demonstrates that a perturbative expansion is sufficient to capture the extensive charging advantage of the SYK$_2$ $x$-battery.\\

Having built up some confidence in the validity of equation \eqref{eq:P_inst_SYK2_pert}, we can now analyze it more closely, to understand the origin of the $L$ scaling.
Each term in the series expansion includes a summation over $O(L)$ terms and, in addition, each term in such a summation is a polynomial, in $L$, of order $O(L^{\frac{i + 1}{2}})$. Here $i$ is the power of $t$ in the corresponding term. Furthermore, each term in the summation has some power in $k$, which takes encodes the size of the corresponding operator appearing in the static Hamiltonian $\hat{H}_0$.
In other words, each summation scales as $O(L^{\frac{i + 1}{2} + 1})$ multiplied by a power of the size, $k$.
Now, keeping in mind that $\mu(\hat{H}_1)$ scales like $O(L)$, the crucial observation is that such a power in $L$ can be \textit{completely} cancelled by each denominator, leaving a factor of order $\left(\frac tL\right)^{-i}$, once again multiplied by size-dependent factors.
From this analysis, we learn that the total $L$ power in each term in the series can be completely reabsorbed in a redefinition of the time variable $t$, subsequently leading to a power which is \textit{independent} of any explicit dependence on $L$.
The full $L$ dependence is thus carried by the \textit{size} of the static Hamiltonian which, crucially in the $x$-battery, scales extensively with the system size $L$. 
Also captured in the same argument is the fact that no extensive quantum advantage exists for the case of the $z$-battery, since in this setup the size of the operators forming $\hat{H}_0$ do not scale with $L$ (being limited to be equal to $2$).

\subsection{Extension to the SYK$_q$ case}

The previous perturbative computation can, in principle, also be extended to the general case of SYK$_q$ Hamiltonians, with $q \geq 4$. 
However, an explicit calculation of the nested commutators appearing in \eqref{eq:P_inst_gen_massaged} is a far more formidable task.
On the other hand, it is  straightforward to compute the $L$-scaling of the terms in order to understand the presence (or absence) of any extensive quantum advantage.

To this end, let us consider the case of generic $q$ and a nested commutator, of order $m$, with even $m = 2 n$.
It follows from \eqref{eq:P_inst_gen_massaged} that such a commutator is accompanied by a factor of order $O(L^{- m})$ (coming from the term $\mu(\hat H_1)^m$ in the denominator and remembering that for $q \geq 4$ the bandwidth of the SYK Hamiltonian scales extensively, \cite{PhysRevD.94.126010}) together with the term $t^{m -1}$.
The nested commutators contribute a total of $m$ copies of the coupling constants $J_{i_1 , \dots , i_q}$ which, again as a result of the average over the Gaussian distributions, must have identical indices in pairs.
From this, and utilizing the usual relation, $\overline{J_{i_1 , \dots, i_q}^2} \propto L^{-(q - 1)}$ for the SYK model, it follows that the average over the Gaussian couplings brings an extra factor of order $O(L^{-n(q - 1)})$ in $L$ which is completely cancelled by summing over the $q - 1$ free indices appearing in each of the $n$ independent coupling constants $J_{i_1 \dots i_q}$. To illustrate, consider the case of four nested commutators and $q=4$. Up to permutations, we need to evaluate terms of the form
\begin{equation}
    \label{eq:nested_commutators_example}
    \frac{1}{L^{2 (4 - 1)}} \sum_{i_2 , i_3 , i_4 , j_2 , j_3 , j_4} J_{1 , i_2 , i_3 , i_4} J_{1 , i_2 , i_3 , i_4} J_{i_2 , j_2 , j_3 , j_4} J_{i_2 , j_2 , j_3 , j_4} \propto O(1) \ ,
\end{equation}
where the index $1$, which is not summed over, represents the index of the initial operator of $\hat{H}_0$ which began the nested commutator. In total then, we have $O(L)$ of these terms since $\hat{H}_0$ is made of $L$ terms.\\

Collecting all these pieces, we conclude that in total we get a factor of order $(\frac tL)^{m - 1}$ (upto possible factors coming from the operator size) which can be reabsorbed into a redefinition of the time variable $t$.
In conclusion, this proves that an extensive quantum advantage requires the presence of large size operators in $\hat{H}_0$ for SYK$_q$ Hamiltonians, and in particular for the $q = 4$ case considered in \cite{Rossini_Rosa_PRL}. The same result can be proven, in the qubit language, as a consequence of the theorem proven in \cite{KRS_advantage}. 

\subsection{Relating the charging power to K-complexity}

The argument above makes manifest the intimate relationship between the average charging power of a given quantum battery on the one hand, and the presence of large operators together with the graph properties of the underlying quench Hamiltonian, on the other. 
As discussed in the main text, the latter two ingredients are the really main actors in the study of operator delocalization dynamics and K-complexity in absence of scrambling.
Hence, it is natural to conjecture a quantitative relation between the notion of K-complexity and the charging power of quantum batteries.
In the main text, just such a relation is made explicit in equation \eqref{eq:P_av_complexity}, for the case of the SYK-like quantum batteries.
We now want to provide a proof of that equation.

To this end, let's assume a quench Hamiltonian, $\hat{H}_1$ which depends on the usual Gaussian couplings, but is defined on a \textit{fixed} graph. As already discussed, the average charging power reads
\begin{equation}
    \label{eq:average_power_SM}
    P_{\mathrm{av}} (t) =  \frac{\overline{\bra{\psi(t)}  \hat{H}_0  \ket{\psi(t)}} - \bra{0}  \hat{H}_0  \ket{0}}{t} =  \frac{\bra{0}  \overline{\hat{H}_0 (t)}  \ket{0} - \bra{0}  \hat{H}_0  \ket{0}}{t} \ ,
\end{equation}
where we average over the Gaussian couplings and have made use of the fact that the ground state $\ket{0}$ is coupling-independent. We will also  assume, without loss of generality, that the static Hamiltonian is traceless. This last assumption means that the ground state energy must be negative, $\bra{0}  \hat{H}_0  \ket{0} < 0$. With this in place, the \textit{averaged} Krylov expansion of $\hat H_0$ is given by
\begin{equation}
    \label{eq:average_Krylov}
    \overline{\hat H_0(t)}  = \sum_{n = 0}^{K - 1} \mathrm{i}^n \overline{\varphi_n(t)} \, \overline{\hat{\mathcal O}_n} \ ,
\end{equation}
where the zeroth-order operator, $\overline{\hat{\mathcal O}_0}$, is actually average-independent and equal to the static Hamiltonian, $\overline{\hat{\mathcal O}_0} = \hat H_0$. 
On the other hand, taking a BCH-like expansion of the static Hamiltonian, and remembering that, as a result of the Gaussian averaging, all couplings must appear in pairs, we can conclude that the \textit{averaged} time-evolved static Hamiltonian satisfies $\overline{\hat H_0(t)} \propto \hat H_0(t)$. Plugging this result into equation \eqref{eq:average_Krylov}, we conclude that $\overline{\hat{\mathcal O}_n} = 0$ for all $n \geq 1$, from which we find
\begin{equation}
    \overline{\hat H_0(t)} = \overline{\varphi_0(t)} \hat H_0(t) \ ,
\end{equation}
which in turns implies that
\begin{equation}
    P_{\mathrm{av}} (t) = \frac{\overline{\varphi_0(t)} - 1}{t} \bra{0}  \hat{H}_0  \ket{0} \ .
\end{equation}

%%%%%%%%%%%%%%%%%%%%%%%%%%%%%%%%%%%%%%%%%%%%%%%%%%%%%%%%%%%%%%%%%%%%%%
\section{An analytic case-study: the star graph}
\label{sec:star_graph}
%%%%%%%%%%%%%%%%%%%%%%%%%%%%%%%%%%%%%%%%%%%%%%%%%%%%%%%%%%%%%%%%%%%%%%
In the star graph all the vertices are connected to a single central or star vertex, with no other connections between nodes. The Hamiltonian for a system of Majorana fermions living on such a network reads
\begin{equation}
    \hat{H} = \mathrm{i} \sum_{i = 1}^{L-1} J_i ~ \hat{\gamma}^i \hat{\gamma}^{L}, \label{eq:star_graph_H}
\end{equation}
where $i = L$ is singled out to be the star vertex and the couplings $J_i$ -- having vanishing mean and variance $\overline{J_i^2} = 1/2$ -- lie on the edges from vertex $i$ to vertex $L$.

% We can compute the time evolution of any operator of size $s$, $\hat{\mathcal{O}}^{(s)}$ by computing the nested commutators $[\hat{H}, \hat{\mathcal{O}}^{(s)}]$, $[\hat{H}, [\hat{H}, \hat{\mathcal{O}}^{(s)}]]$, etc. In practice, however, it is more convenient to compute the Krylov operators and Krylov coefficients $\{\hat{\mathcal{O}}^{(s)}_n, b^{(s)}_n\}$, where $0 \leq n \leq K-1$, $n=0$ is our initial operator and $K$ is the dimensionality of Krylov space. Using this formalism, the time-evolved operator becomes
% \begin{equation}
%     \hat{\mathcal{O}}^{(s)}(t) = \sum_{n = 0}^{K-1} i^n \varphi^{(s)}_n(t) \hat{\mathcal{O}}^{(s)}_n,
% \end{equation}
% where $\varphi^{(s)}_n(t)$ are wavefunctions that satisfy the differential equations
% \begin{equation}
%     \dot{\varphi}^{(s)}_n(t) = b^{(s)}_n \varphi^{(s)}_{n-1}(t) - b^{(s)}_{n+1} \varphi^{(s)}_{n+1}(t). %\label{eq:phi_diff_eqns}
% \end{equation}
% Hence, knowledge of the Krylov coefficients $b^{(s)}_n$ is enough to solve these differential equations.

Since Krylov space and all its related quantities is highly dependent on the initial operator that we choose, we will first study the simpler dynamics of the operator
\begin{equation}
    \hat{\mathcal{O}}_0^{(s)} = 2^{s/2} \prod_{i=1}^s \hat{\gamma}^i, \label{eq:initial_op_s}
\end{equation}
before we move to the dynamics of the $x-$ and $z-$ batteries, which will have as initial operators a linear combination of the former.

It turns out that the dimension $K$ of the Krylov space associated to this graph and initial operator as in quation \eqref{eq:initial_op_s} is $K=3$ for all values $s < L$. This can be easily seen by taking the simplest initial operator $\hat{\mathcal{O}}_0 \sim \hat{\gamma}^1$. $\hat{\mathcal{O}}_1$ is $[\hat{H}, \hat{\mathcal{O}}_0] \sim \hat{\gamma}^{L}$, which may be interpreted as the first nested commutator representing a jump from vertex $i = 1$ to vertex $i = L$ (the star vertex). Since the star vertex is connected to all other vertices, the next jump proceeds to $\sim \sum_{i=1}^{L-1} \hat{\gamma}^i$, but since $\hat{\gamma}^1$ was already visited, we end up with $\hat{\mathcal{O}}_2 \sim \sum_{i=2}^{L-1} \hat{\gamma}^i$. At this point, with all vertices having been visited, the next nested commutator yields $\hat{\mathcal{O}}_3 = 0$. One encounters a similar situation when starting with an initial operator of size different than unity. The equations for the Krylov wavefunctions, equation \eqref{eq:phi_diff_eqns} in the main text, can be exactly solved in the case of the star graph, giving
\begin{eqnarray}
    \varphi_0(t) &=& \frac{b_2^2 + b_1^2 \cos{\left( \sqrt{b_1^2 + b_2^2}~t\right)}}{b_1^2 + b_2^2}\ , \nonumber \\
    \varphi_1(t) &=& \frac{b_1 \sin \left( \sqrt{b_1^2 + b_2^2} ~t \right)}{\sqrt{b_1^2 + b_2^2}} \ , \label{eq:phis_star}\\
    \varphi_2(t) &=& \frac{b_1 b_2 \left( 1 - \cos \left(\sqrt{b_1^2 + b_2^2}~t \right) \right)}{b_1^2 + b_2^2}.\nonumber
\end{eqnarray}
Moreover, the Lanczos coefficients $b_1, b_2$ can also be written as analytic functions of the couplings $J_i$. Specifically,
\begin{equation}
    b_1(s) = \sqrt{\sum_{i = 1}^s J_i^2}, \quad b_2 (s) = \sqrt{\sum_{i = s+1}^{L-1} J_i^2}. \label{eq:b1_b2_star_o_s}
\end{equation}
However, notice that the functional form of the Krylov wavefunctions, euqations \eqref{eq:phis_star}, are \textit{independent} of the initial operator that we choose on the graph as long as $K=3$. On the other hand, the form of the Lanczos coefficients, equation \eqref{eq:b1_b2_star_o_s}, do depend on the initial operator, in this case labelled by $s$ in \eqref{eq:initial_op_s}. Note also that $b_1^2 + b_2^2 = \sum_{i = 1}^{L-1} J_i^2 = \mu^2$, where $\mu$ is the bandwith of Hamiltonian \eqref{eq:star_graph_H}; a fact that can be seen from the eigenvalue properties of Majorana gamma matrices. Since we are interested in taking ensemble averages, computations can be made more efficient by noticing the following relations
\begin{eqnarray}
    \overline{F[b_2]}(s) &=& \overline{F[b_1]}(L-1-s) \ , \\
    \overline{F[\mu]} &=& \overline{F[b_1]}(L-1) \ ,
\end{eqnarray}
among arbitrary functions of the Lanczos coefficients. In particular, this means that $\overline{b_2^2}(s) = \overline{b_1^2}(L-1-s)$ and $\overline{\mu^2} = \overline{b_1^2}(L-1)$, so we need only compute $\overline{b_1^2}$.
% However, notice that unlike the $J_i$ couplings, $b_1$, $b_2$ and $\mu$ don't have a vanishing mean. These are,
% \begin{equation}
%     \overline{b_1}(s) = \frac{\Gamma\left( \frac{s+1}{2} \right)}{\Gamma\left( \frac{s}{2} \right)}, \quad \overline{b_2}(s) = \overline{b_1}(L-1-s), \quad \overline{\mu} = \overline{b_1}(L-1)
% \end{equation}
% where $\Gamma(z) = \int_0^{+\infty} dx~ x^{z-1} e^{-x}$ is the standard Gamma function.

As seen in equation \eqref{eq:P_av_complexity} in the main text, the average power is essentially determined by $\overline{\varphi_0}(t)$. However, in order to check whether there is an \textit{extensive quantum charging advantage} it is important to rescale the quench Hamiltonian by its bandwidth. This results in a Hamiltonian whose bandwidth is equal to one and new rescaled couplings $J_i \to J_i/\mu$ and leads to the identity $b_1^2 + b_2^2 = 1$ for the rescaled Lanczos coefficients, which in turn makes it convenient to define $b \equiv b_1$, while $b_2^2 = 1 - b^2$. The resulting statistical properties of our system can be completely characterized by the statistical moments of $b$,
\begin{equation}
    \overline{b^{2n}}(s) = \frac{\prod_{j=0}^{n-1} (s + 2j)}{\prod_{j=0}^{n-1} (L - 1 + 2j)}, \label{eq:b_moments}
\end{equation}
with $n\in\mathbf{N}$, a natural number. Odd moments of $b$ can also be computed but are not useful in what follows so we choose to not include them. A particularly interesting case is the first moment, $n=1$, which gives $\overline{b^2} = s/(L-1)$. This result is actually independent of the probability distribution for the couplings $J_i$ and is true for independent and identically distributed random variables.

The rescaled wavefunctions also receive important simplifications, which in the case of interest for the charging power, $\varphi_0 = 1 + b^2 (\cos(t) - 1)$. Then,
\begin{equation}
    \overline{\varphi_0}(t) = 1 + \frac{s}{L-1} ( \cos(t) - 1),
\end{equation}
which also holds for any kind of random variables.

\begin{figure}[t!]
\begin{center}
    \includegraphics[width=.5\hsize]{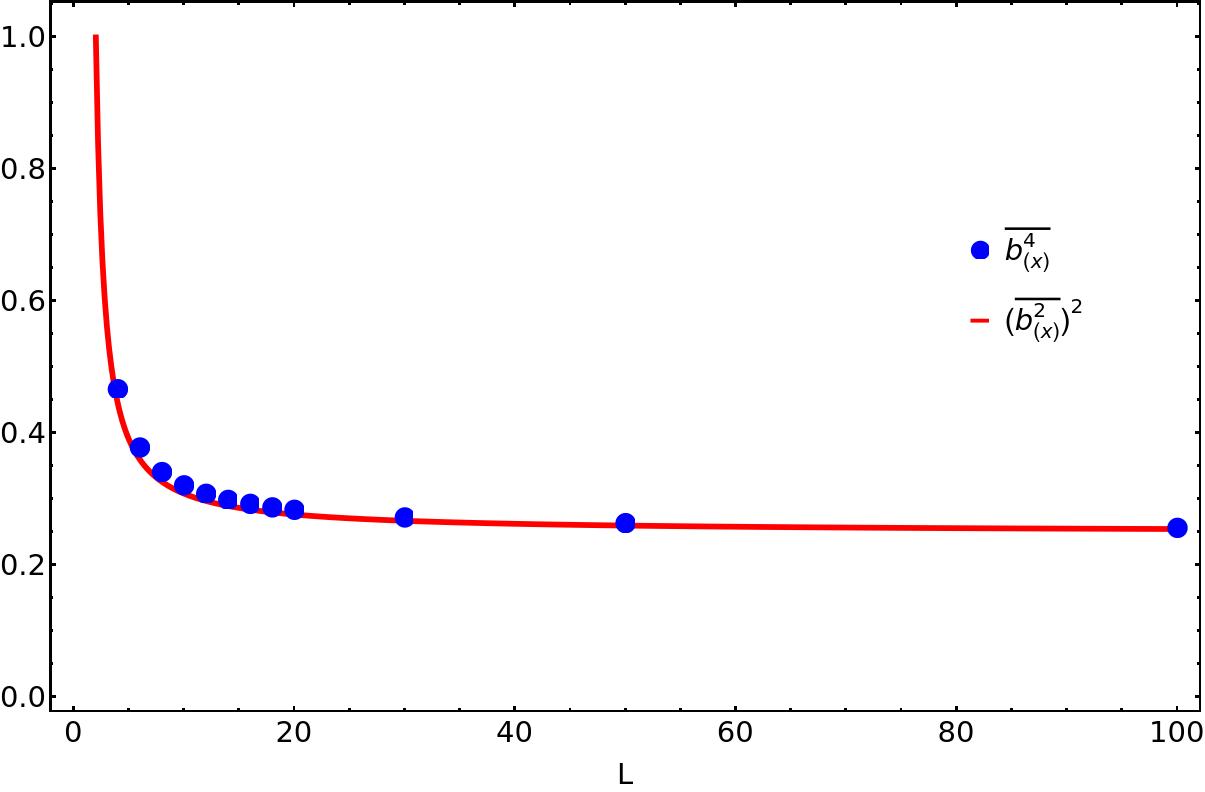}\\
    \caption{Comparison between the numerically computed $\overline{b^4_{(x)}}$ and $\left(\overline{b^2_{(x)}}\right)^2 = (1 - (L-2)/(2(L-1)))^2$, as functions of $L$.}
\label{fig:b4_numerics_vs_b2_2}
\end{center}
\end{figure}

We may also compute the average Krylov Complexity for the rescaled Hamiltonian,
\begin{equation}
    \overline{C_K}(t) = \frac{s}{L-1}~\sin^2(t) + \frac{2 s (L - 1 - s)}{(L-1)(L+1)}~(\cos(t) - 1)^2, \label{eq:CK_star_averaged_trig}
\end{equation}
or in terms of the charging power,
\begin{equation}
    \overline{C_K}(t) = \frac{s}{L-1}~\sin^2(t) + \frac{2 (L-1) (L - 1 - s)}{s (L+1)}~(\overline{\varphi}_0(t) - 1)^2. \label{eq:CK_star_averaged_pow}
\end{equation}
Let's finally study the charging power for the $x-$ and $z-$batteries. This involves choosing as initial operators,
\begin{eqnarray}
    \hat{\mathcal{O}}_0^{(x)} &=& \sqrt{\frac{2}{L}}~\sum_{j=1}^{L/2}\left(\sqrt{2}\right)^{2j-1} (-\mathrm{i})^{j-1} \prod_{p=1}^{2j-1} \hat{\gamma}^p \ , \label{eq:initial_op_x} \\
    \hat{\mathcal{O}}_0^{(z)} &=& \sqrt{\frac{2}{L}}(-2 \mathrm{i})~\sum_{j=1}^{L/2} \hat{\gamma}^{2j-1} \hat{\gamma}^{2j} \ ,
\end{eqnarray}
respectively. Beginning with the $x-$battery; schematically, in terms of the operators in \eqref{eq:initial_op_s}, $\hat{\mathcal{O}}_0^{(x)} \sim \sum_{j=1}^{L/2} \hat{\mathcal{O}}_0^{(s = 2j-1)}$, which then evolves in time as $\hat{\mathcal{O}}_0^{(x)}(t) \sim \sum_{j=1}^{L/2} \hat{\mathcal{O}}_0^{(s = 2j-1)}(t) = \sum_{n=0}^{2} \mathrm{i}^n \sum_{j=1}^{L/2} \varphi_n^{(s)}(t) \hat{\mathcal{O}}_n^{(s=2j-1)}$. On the other hand, it can also be expanded in its Krylov basis as $\hat{\mathcal{O}}_0^{(x)}(t) = \sum_{n=0}^{K-1} \mathrm{i}^n \varphi_n^{(x)}(t) \hat{\mathcal{O}}^{(x)}_n$. Both are valid expressions for the time evolution of $\hat{\mathcal{O}}_0^{(x)}$, but we want to stress that it is in general not trivial to relate Krylov operators and Krylov wavefunctions of two different initial operators, even in the case where they are related by linear combination, as in \eqref{eq:initial_op_x}. The corresponding Lanczos coefficient $b_1$ now becomes
\begin{equation}
    b_1^{(x)} = \sqrt{J_1^2 + \sum_{i=1}^{L/2-1} \left( 1 - \frac{2 i}{L}\right) \left( J^2_{2i} + J^2_{2i+1}\right)} \ ,
\end{equation}
while the relation $b_1^2 + b_2^2 = \mu^2$ still holds here. Thus, when we rescale the Hamiltonian, we will have, again, $b \equiv b_1$ and $b_2^2 = 1 - b^2$. The average charging power for the $x-$battery is then
\begin{equation}
    P_{\text{av}}^{(x)}(t) = \frac{L}{2} \left( 1 - \frac{1}{2} \frac{L-2}{L-1} \right) \frac{1 - \cos(t)}{t}, \label{eq:star_average_power}
\end{equation}
while the maximum charging power is given by
\begin{equation}
    P_{\text{max}}^{(x)}(L) = \frac{0.724611}{2} \left( 1 - \frac{1}{2} \frac{L-2}{L-1} \right) L,
\end{equation}
which manifestly shows an \textit{extensive} scaling, as result of both large operators and a highly connected underlying graph. The average Krylov complexity is
\begin{equation}
    \overline{C_K^{(x)}}(t) = \left( 1 - \frac{1}{2} \frac{L-2}{L-1} \right) \left[ \sin^2(t) + 2(1 - \cos(t))^2 \right] -2\overline{b^4_{(x)}} (1 - \cos(t))^2,
\end{equation}
where $\overline{b^4_{(x)}}$ is a complicated function of $L$ that nonetheless can both be easily computed numerically for all $L$ and whose analytic expression can be in principle obtained through repeated use of equation \eqref{eq:b_moments}. Moreover, its asymptotics are $\overline{b^4_{(x)}}(L=2) = 1$ and $\overline{b^4_{(x)}}(L \to \infty) = 1/4$. Numerically, its values are not far away from $\left(\overline{b^2_{(x)}}\right)^2 = (1 - (L-2)/(2(L-1)))^2$ (see figure \ref{fig:b4_numerics_vs_b2_2}), so we may use this expression to give a final analytic expression in terms of the average charging power, equation \eqref{eq:star_average_power},
\begin{equation}
    \overline{C_K^{(x)}}(t) \approx \left( 1 - \frac{1}{2} \frac{L-2}{L-1} \right) \sin^2(t) + 8 \left( 3 - \frac{4}{L} \right) \frac{t^2 P_{\text{av}}^2}{L^2}.
\end{equation}

Finally, let's discuss the $z-$battery following the same steps as for the case of the $x-$battery case. Here we find the Lanczos coefficient
\begin{equation}
    b_1^{(z)} = \sqrt{\frac{2}{L}} \sqrt{\sum_{i=1}^{L-2} J_i^2},
\end{equation}
while $b_2^{(z)}$ is again related to $b_1^{(z)}$ through the bandwith of the star Hamiltonian. Interestingly, $b_1^{(z)}$ is essentially $b_1^{(s=L-2)}$ (equation \eqref{eq:b1_b2_star_o_s}) albeit with an extra $\sqrt{L}$ suppression, which is essentially responsible for eliminating the quantum charging advantage in the $z-$battery, giving
\begin{equation}
    P_{\text{av}}^{(z)}(t) = \frac{L-2}{L-1} ~ \frac{1 - \cos(t)}{t},
\end{equation}
and maximum charging power
\begin{equation}
    P_{\text{max}}^{(z)}(L) = 0.724611 ~ \frac{L-2}{L-1}.
\end{equation}
Finally, the average Krylov complexity,
\begin{equation}
    \overline{C_K^{(z)}}(t) = \frac{2}{L} \left( \frac{L-2}{L-1} \sin^2(t) + 2~ \frac{L-2}{L-1}(1 - \cos(t))^2 \right),
\end{equation}
can be expressed in terms of the average charging power as,
\begin{equation}
    \overline{C_K^{(z)}}(t) = \frac{2}{L} \left( \frac{L-2}{L-1} \sin^2(t) + 2~  \frac{(L-1)^2}{(L-2)(L+1)}~ t^2 P_{\text{av}}^2 \right).
\end{equation}

% Finally, we give the ratio $R(t)$, defined as the ratio between $C_K(t)$ for operators of size $L/2$ and operators of size $1$. It is
% \begin{equation}
%     \overline{R}(t) = \frac{L}{2}~\frac{2L - 1 + 3 \cos(t)}{3(L-1) - (L-5) \cos(t)}
% \end{equation}
% and so it's not very time-independent. But this makes sense since, contrary to the other cases, we have a Krylov space with small $K$. This, in turn, also tells you that the time-independence observed in the other cases arises from the "superposition" of a large number of "modes". Here, having only 3 "modes" makes that any function that you try to build out of them will necessarily be quite periodic. In any case, the maximum value $\overline{R_{\text{max}}}(t) = L/2$, while the minimum one is $\overline{R_{\text{min}}}(t) = L/4$. Lastly, we can average this ratio over one period to obtain
% \begin{equation}
%     \overline{R}(L) = \frac{\sqrt{2(L+1)(L-2)} - 6}{L-5}~\frac{L}{4},
% \end{equation}
% which shows a linear growth with $L$, $\overline{R}(L \gg 1) \approx 2^{-3/2}~L$.

\end{appendices}

\end{document}